\documentclass{aa}
\usepackage{graphics}
\def\erre{{r}}

\def\ng{n_G}
\def\nr{n_R}

\def\hmpc{\rm \,h^{-1}\,Mpc}
\def\etal{{\rm et al.}\ }
\def\x{$\xi(r)$\ }
\def\xr{$\xi(r)$\ }
\def\xs{$\xi(s)$\ }
\def\xir{$\xi(r)$\ }
\def\xis{$\xi(s)$\ }
\def\xip{$\xi(r_p,\pi)$\ }
\def\xip{$\xi(r_p,\pi)$\ }
\def\wp{$w_p(r_p)$\ }
\def\sg{\sigma_{12}(1)}
\def\kms{\,{\rm km\,s^{-1}}}

\def\vva{{\bf v}_1\ }
\def\vvb{{\bf v}_2\ }

\def\n_med{{\left<n\right>}}

\def\begc{\begin{center} }
\def\endc{\end{center} } 
\def\begf{\begin{figure} }
\def\endf{\end{figure} }

\def\j3{{J_3}}
\def\Dmat{{\bf D}}
\def\Cmat{{\bf C}}
\def\Tmat{{\bf T}}

\def\fig #1, #2, #3 {
\smallskip
\centerline{\psfig{figure=#1,height=#2 in,width=#3 in}}
}

\begin{document}

\thesaurus{06         
              (03.11.1;  
               16.06.1;  
               19.06.1;  
               19.37.1;  
               19.53.1;  
               19.63.1)  
             }
\title{The ESO Slice Project (ESP) Galaxy Redshift Survey 
\thanks{Based on observations collected at the European Southern
Observatory, La Silla, Chile.} }
\subtitle{VII. The Redshift and Real--Space Correlation Functions}


\author{
L. Guzzo\inst{1}
\and
J.G. Bartlett\inst{2}
\and
A. Cappi\inst{3}
\and
S. Maurogordato\inst{4}
\and
E. Zucca\inst{3,5}
\and
G. Zamorani\inst{3,5}
\and
C. Balkowski\inst{6}
\and
A. Blanchard\inst{2}
\and
V. Cayatte\inst{6}
\and
G. Chincarini\inst{1,7}
\and
C.A. Collins\inst{8}
\and
D. Maccagni\inst{9}
\and
H. MacGillivray\inst{10}
\and
R. Merighi\inst{3}
\and
M. Mignoli\inst{3}
\and
D. Proust\inst{6}
\and
M. Ramella\inst{11}
\and
R. Scaramella\inst{12}
\and
G.M. Stirpe\inst{3}
\and
G. Vettolani\inst{5}
}

\offprints{L. Guzzo, guzzo@merate.mi.astro.it}

\institute{ 
Osservatorio Astronomico di Brera, 
via Bianchi 46, 23807  Merate (LC), Italy
\and
Observatoire Astronomique, Universit\'e Luis Pasteur,
11 rue de l'Universit\'e, 67000 Strasbourg, France,\\ 
(unit\'e associ\`e au CNRS -- UMR7550) 
\and
Osservatorio Astronomico di Bologna, 
via Zamboni 33, 40126 Bologna, Italy
\and
CERGA, Observatoire de la C\^ote d'Azur, 06304 Nice Cedex 4, France
\and
Istituto di Radioastronomia del CNR, 
via Gobetti 101, 40129 Bologna, Italy
\and
Observatoire de Paris, DAEC, 5 Pl. J.Janssen, 92195 Meudon, France
\and
Dipartimento di Fisica, Universit\`a degli Studi di Milano, 
via Celoria 16, 20133 Milano, Italy
\and
School of Engineering, Liverpool John Moores University, 
Byrom Street, Liverpool L3 3AF, United Kingdom
\and
Istituto di Fisica Cosmica e Tecnologie Relative, 
via Bassini 15, 20133 Milano, Italy
\and
Royal Observatory Edinburgh, 
Blackford Hill, Edinburgh EH9 3HJ, United Kingdom
\and
Osservatorio Astronomico di Trieste, 
via Tiepolo 11, 34131 Trieste, Italy
\and
Osservatorio Astronomico di Roma, 
via Osservatorio 2, 00040 Monteporzio Catone (RM), Italy
}

\date{Received 00 - 00 - 0000; accepted 00 - 00 - 0000}

\maketitle

\markboth {L.~Guzzo et al.: 
Clustering in the ESP Galaxy Redshift Survey}{}

\begin{abstract}
We present analyses of the two-point correlation properties of the 
ESO Slice Project (ESP) galaxy redshift survey, both in redshift 
and real space.  From the redshift--space correlation function \xis
we are able to trace positive clustering out to
separations as large as $ 50\hmpc$, after which $\xi(s)$ smoothly breaks 
down, crossing the zero value between 60 and $80\hmpc$.  This is best 
seen from the whole magnitude--limited redshift catalogue, 
using the $J_3$ minimum--variance weighting estimator.  
\xis is reasonably well described by a shallow power law with 
$\gamma\sim 1.5$ between 3 and $50\hmpc$, while on smaller scales 
($0.2-2\hmpc$) it has a shallower slope ($\gamma\sim 1$).  
This flattening is shown to be mostly due to the redshift--space damping
produced by virialized structures, and is less evident when volume--limited
samples of the survey are analysed.  

We examine the full effect of 
redshift--space distortions by computing the two--dimensional 
correlation function \xip, from which we project out the real--space
\xir below $10\hmpc$.  This function is well described by a 
power--law model $(r/r_o)^{-\gamma}$, with $r_o=4.15^{+0.20}_{-0.21} 
h^{-1}\,$ Mpc and $\gamma=1.67^{+0.07}_{-0.09}$. 

Comparison to other redshift surveys shows a consistent picture 
in which galaxy clustering remains positive out to separations of $50 
\hmpc$ or larger, in substantial agreement with the results obtained 
from angular  
surveys like the APM and EDSGC.  Also the shape of the two--point
correlation function is remarkably unanimous among these data sets, 
in all cases requiring more power above $5\hmpc$ (a `shoulder'), 
than a simple extrapolation of the canonical \xir=$(r/5)^{-1.8}$.

The analysis of \xis for volume--limited subsamples with different luminosity
shows evidence of luminosity segregation only for the most luminous sample
with $M_{b_J}\le -20.5$.  For these galaxies, the amplitude of clustering
is on all scales $>4\hmpc$ about a factor of 2 above that of all 
other subsamples containing less luminous galaxies.  When redshift--space 
distortions are removed through projection of \xip, however, a weak dependence
on luminosity is seen at small separations also at fainter magnitudes, 
resulting in a growth of $r_o$ from $3.45_{-0.30}^{+0.21}\hmpc$ to 
$5.15_{-0.44}^{+0.39}\hmpc$, when the limiting absolute magnitude
of the sample changes from $M=-18.5$ to $M=-20$.  This effect is
masked in redshift space, as the mean pairwise velocity dispersion
experiences a parallel increase, basically erasing the effect of 
the clustering growth on \xis.

\end{abstract}

keywords{ -- Cosmology -- Large--Scale Structure}

\section{Introduction}

The spatial two-point correlation function, \x, is probably the most
classical statistic used in cosmology for clustering analyses.  
Since its early applications (e.g. Peebles 1980), obtaining 
reliable estimates of \x on large ($>5-10\hmpc$) scales became one of the 
main statistical motivations for enlarging the available 3D samples
through new, wider and deeper redshift surveys.  The principal reason for this 
is that on scales 
sufficiently large for the fluctuations to be still in the linear clustering 
regime, the {\it shape} of \x is expected to be preserved during gravitational
clustering growth.  Comparison of observations with 
models is therefore simpler, because in this case the theoretical 
description of clustering does not require the full non-linear gravitational
modelling, which is on the contrary necessary at small separations.
If we are able to measure accurately \x 
[or its Fourier dual, the power spectrum P(k)], on large enough scales, 
we shall have a measure of the distribution of the {\it initial}
fluctuation amplitudes.
In addition, if the initial density field was described by a Gaussian 
statistics, this will be all the statistical information we 
need to completely characterise the initial field itself.  
One further reason for pushing measures of \x to larger and larger scales,
is the possibility to correctly estimate its zero--point, 
i.e. the scale on which correlations become negative.  This is a specific
prediction of any viable model, as it reflects the turnover and peak scale 
in the power spectrum, fingerprint of the horizon size at
the matter--radiation equivalence epoch.  Clearly, as much as the 
weakness of clustering in this regime is a benefit for the theory,
it represents a hard challenge for the observations: statistical 
fluctuations destroy any possibility to detect significant features, 
such as the zero--point of \x, if the sample under study does not contain 
enough objects at comparable separations.   

Historically, following the pioneering works of the mid--seventies
(see Rood 1988 for a review), the industry of redshift measurements 
exploded in the eighties,  
with the completion first of the CfA1 (Davis et al. 1982), then of 
the Perseus-Pisces (Giovanelli et al. 1986) and CfA2/SSRS2 (Geller \& Huchra 
1989; da Costa et al. 1994) surveys.  These first large surveys (containing several $10^3$ redshifts),
produced significant advances in the estimation of \x on small and
intermediate scales 
(e.g. Davis \& Peebles 1983, De Lapparent et al. 1988).
However, they also showed the existence of structures with dimensions 
comparable to their depth.  This, while on one side giving rise 
to speculations about the very existence of a transition to homogeneity on 
large scales (see Guzzo 1997 for a recent review of this problem), explicitly 
indicated the need for larger redshift samples.  In more recent years,
there have been a few significant attempts to fulfil this need, as for
example the surveys based on the IRAS source catalogues (see e.g. Strauss 
1996), the Stromlo--APM survey (Loveday et al. 1992b), and especially the Las
Campanas Redshift Survey (LCRS, Shectman et al. 1996).  Our contribution
along this direction has been the realization of the ESO Slice Project
(ESP) galaxy redshift survey, that was completed between 1993 and 1996
(Vettolani \etal 1997, Vettolani et al. 1998, V98 hereafter).  The technical aims of the ESP survey were 
to exploit on one side the availability of new deep photometric galaxy 
catalogues (in our case 
the EDSGC, \cite{edsgc}), and on the other the multiplexing 
performances of fibre spectrographs, as in the specific case of the 
Optopus fibre coupler available at ESO (Avila et al. 1989).  
The main scientific goals were to estimate the galaxy luminosity 
function over a large 
and homogeneously selected sample with a large dynamic range in magnitudes 
(see Zucca et al. 1997, paper II hereafter), and to measure the 
clustering of galaxies over a hopefully fair sample of the Universe.   
The ESP survey, during about 25 nights of observations, produced a $\sim 85
\% $ complete sample 
of 3342 galaxies with reliable redshift.  Its combination of depth and 
angular extension is paralleled at present only by the Las Campanas 
Redshift Survey (LCRS, Shectman et al. 1996), which has the advantage
of containing  a larger number of redshifts.  One main difference of the
LCRS with respect to the ESP, is that it is selected in the red, which
makes comparison with the results presented here particularly interesting.
One important advantage of the ESP is that it is a purely magnitude--limited
sample, which makes modelling of the selection function easier than in
the case of the LCRS, where an additional surface--brightness selection
was applied.  

In this paper we shall discuss the two-point correlation properties of the ESP 
survey, both in redshift and in real space.  Previous or parallel papers, 
in addition to that describing the above mentioned $b_J$--band luminosity 
function (Zucca \etal 1997), deal with the  
scaling properties (Scaramella \etal 1998), the properties of
groups (Ramella \etal 1998), and potential biases
in the estimate of galaxy redshifts (Cappi \etal 1998).  The survey in
general is described in Vettolani et al. (1997), while the redshift data 
catalogue is presented in V98.  Here we also marginally 
discuss the small--scale galaxy dynamics, for its effects on redshift--space
clustering.  However, a proper discussion of the small--scale pairwise 
velocity dispersion and related topics, will be presented in a separate
paper (Guzzo \etal, in preparation).

The paper is organized as follows.   Section 2 briefly summarises the 
properties of the redshift catalogue.  Section 3 discusses the 
techniques used to estimate two--point correlations and their errors.
Sections 4 and 5 present the redshift-- and real--space correlation 
functions, discussing their dependence on luminosity and comparing
results with other surveys.  In section 6 possible biases deriving 
from the observational setup are investigated.  The main results are
summarised in section 7.

\section{The Survey}\label{sec-data}

	The ESP redshift survey consists of a strip of sky $1^o$ 
thick (declination) by $22^o$ long (right ascension) near the 
South Galactic Pole, plus an additional strip of length $5^o$
situated five degrees to the west of the main strip.  
In total, this covers about 25 square degrees at a mean 
declination of $-40.25^o$ between right ascensions $22^h30^m$ 
and $01^h20^m$.  The target galaxies were selected from the 
Edinburgh--Durham Southern Galaxy Catalogue (EDSGC, Heydon--Dumbleton
et al. 1989), which is complete to $b_J=20.5$.  The angular
correlation properties of this catalogue were studied in detail
in Collins et al. (1992).   The limiting magnitude of the ESP is
$b_J=19.4$, chosen in order to have the best 
match of the number of targets to the number of fibres in the
field of the multi--object spectrograph Optopus, mounted at
the Cassegrain focus of the ESO 3.6~m telescope.
More details on the observing strategy and the properties of the spectroscopic
data are given in Vettolani \etal (1997), and in V98.  

%
\begin{figure*}
\resizebox{\hsize}{!}{\includegraphics{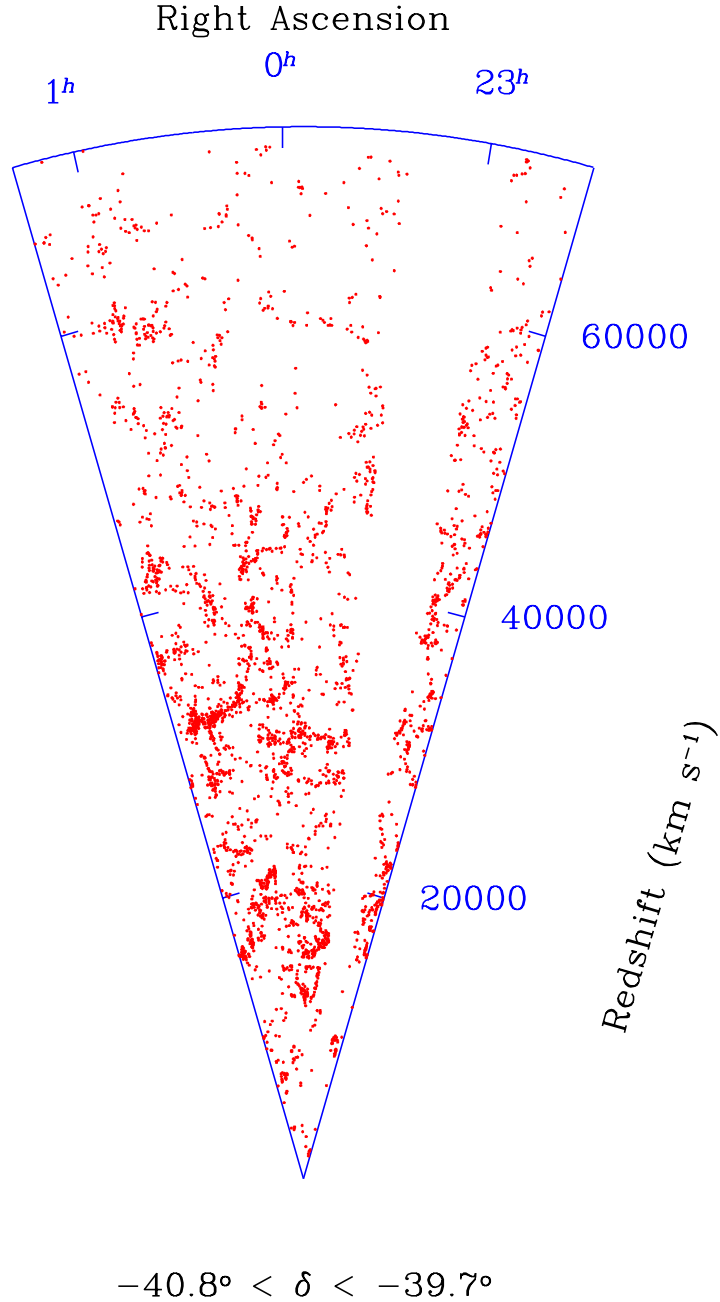}}
\caption{Cone diagram showing the large-scale distribution of galaxies in 
the ESP survey.}
\label{fig-cone}
\end{figure*}
%

The final galaxy redshift catalogue contains 3342 entries, and 
is 85\% redshift--complete within the survey area.   
%
%
At its effective depth ($z\simeq0.16$), the transverse linear 
dimensions of the survey are $\sim 210\hmpc$ by $6.5\hmpc$, while 
its volume is $\sim 1.9 \cdot 10^5{\rm \,h^{-3}\,Mpc^{3}}$.

As can be seen in V98, the survey geometry is fairly complex,
being composed by two rows of circular Optopus fields, partly
overlapping each other.  This results in the presence of interstices
in which galaxy redshifts were not measured, and which obviously have 
to be considered carefully in the analysis of clustering.  
Also, the completeness of each 
Optopus field is not strictly constant (V98).
This will require a weight to
be applied to each object, depending on its parent field (see \S 
\ref{sec-est}).
The median internal error on the redshift measurements is $64 \kms$ 
for absorption--line estimates, and $31 \kms$ for emission--line
estimates (V98).

In Figure~\ref{fig-cone} we show a cone diagram of the galaxy distribution in 
the 
survey.  One first qualitative remark to be made, in view of the 
use we shall do of the data in this paper, is that the visual impression 
from this figure suggests - contrary to more "local" surveys as, e.g. CfA2, a typical size of structures which is smaller than the survey dimensions.  This gives us hope that, possibly, clustering measures 
extracted from this survey would be sufficiently representative of the general
properties of our Universe.  Clearly, the volume covered by the ESP is still 
small, due to its limited area coverage on the sky, yet the linear 
sampling of large-scale structures in two dimensions is unprecedented,
and matched only by the even larger LCRS.  

Preliminary to the analysis, we performed a series 
of necessary corrections to the raw data catalogue.  We first applied 
K-corrections to the observed $b_J$ apparent magnitudes in the same way as 
discussed in detail in Zucca \etal (1997). Given the location of the ESP area 
in the region of the South Galactic Pole, we did not apply any correction for 
galactic extinction to the observed magnitudes. 
Observed heliocentric radial velocities were converted to the reference frame 
of the Local Group using the transformation of Yahil \etal (1977).  (We also 
checked that correcting velocities to the CMB reference frame does not 
produce any difference in the results). 
We then computed comoving, luminosity, and angular diameter distances for 
each galaxy, adopting a cosmological model with H$_o = 100$ h $\kms$ 
Mpc$^{-1}$, and q$_o=0.5$.  

The correlation analyses are first performed on the whole 
apparent--magnitude limited catalogue.  To do this, we apply the minimum 
variance technique (see \S~\ref{sec-est}), for which we use the
survey selection function as computed in Zucca \etal (1997).  
We also exclude from 
this analysis those galaxies lying outside of the range of comoving 
distances $100\le D_{com}\le 500 \hmpc$, 
to avoid using those regions of the sample where respectively 
either the survey volume or the selection function are dangerously small. 
The sample trimmed in this way contains 2850 galaxies, with a minimum
luminosity corresponding to $M_{b_J}=-15.6+5\,{\rm log\, h}$.
Using the whole magnitude-limited sample is an attempt to extract the 
maximum signal from the available data, but can have some drawbacks, as
the contribution of galaxies with different luminosities is not homogeneous
over the sampled scales.   

Therefore, to evidence possible biases and study the general behaviour of 
clustering with luminosity, we also construct a set of volume--limited 
subsamples, defined in Table 1.  Each column lists, 
respectively, 
(1) absolute magnitude limit, 
(2) luminosity distance computed from the magnitude limit 
taking into account the K-correction, 
(3) corresponding comoving distance limit, 
(4) redshift limit, 
(5) lower cut in comoving distance,
(6) effective volume of the sample between the two distance boundaries,  
(7) total number of galaxies.   Given the small thickness of the 
ESP slice, the lower distance limits are a safeguard to 
exclude that part of the samples where galaxy density is potentially
undersampled by bright galaxies, thus introducing shot noise.
These low cut--off values have been computed as the distance within which,
for that absolute magnitude limit, one expects less than 10 galaxies 
within the corresponding ESP volume, on the basis of the ESP luminosity
function.
The main analyses we shall discuss in the following are based on the
first four samples in the table.  These offer the best compromise 
between having enough volume sampled and a good statistics.  However, we
also select a more ``extreme'' sample with $M\le-20.5$, which contains
only 292 galaxies, to study in more detail the existence of luminosity
segregation at high luminosities.

\begin{table*}
\label{tab-vlim}
\begin{flushleft}
\begin{tabular}{rrrrrcc}
\hline
 $M_{lim}$ & $D_{lum(max)}$ & $D_{com(max)}$ & z$_{max}$ & $D_{com(min)}$ & $Vol$& $N_{gal}$ \\
\hline
%
 $-18.5$ & 328.5 & 296.8 & 0.106874 & 65  & $6.109 \cdot 10^4$ & 823 \\
 $-19.0$ & 398.7 & 353.1 & 0.129082 & 80  & $1.027 \cdot 10^5$ & 924 \\
 $-19.5$ & 483.4 & 418.3 & 0.155607 & 100 & $1.705 \cdot 10^5$ & 819 \\
 $-20.0$ & 590.3 & 496.6 & 0.188737 & 135 & $2.834 \cdot 10^5$ & 521 \\
 $-20.5$ & 738.4 & 598.4 & 0.234030 & 200 & $4.871 \cdot 10^5$ & 292 \\
\hline
\end{tabular}
\caption[]{Properties of the volume--limited subsamples analysed.  
Distances (in Mpc), volumes (in Mpc$^3$) and absolute magnitudes 
($b_J$ system), are computed for h=1, $q_\circ=0.5$}
\end{flushleft}
\end{table*}


\section{Computing the Two-Point Correlation Function}

\subsection{Estimators}
\label{sec-est}

There is a long history and literature concerning the theory
of statistical estimators of the two--point correlation function $\xi(r)$.
The most widely used one (Davis \& Peebles 1983), is given by
\begin{equation}
\label{est1}
\xi(r) = \frac{2\nr}{\ng} \frac{GG(r)}{GR(r)} - 1\,\,\,\, ,
\end{equation}
where $\ng$ and $\nr$ are the mean space densities 
(or equivalently the total numbers), of objects in the two catalogues, $GG(r)$ represents the number of {\it independent} Galaxy-Galaxy 
pairs with separation between $r$ and $r+dr$, while $GR(r)$ 
is the number of Galaxy-Random pairs, computed with respect to a random 
catalogue of points distributed with the same redshift selection function 
and the same geometry of the real sample (in other words, a Poisson 
realization of the catalogue). 
In our case, this implies reproducing the exact distribution of 
Optopus fields on the sky, and applying the same selection process as done 
on the real data.  One example is
the "drilling" of some areas as it was originally done on the EDSGC 
around the position of very bright stars (\cite{edsgc})\footnote{Within the ESP 
area there are about 10 drill holes 
with a typical diameter of 0.2 deg.}.  

The number of $GG(r)$ or $GR(r)$ pairs in eq.~\ref{est1} can be formed 
in general including an arbitrary statistical weight, as  $GG(r) = 
\sum_{(i,j)} w_{ij}(\erre) $, that can depend both on the distances of
the two objects from the observer ($d_i$ and $d_j$), and their separation
($r=|{\bf d_i - d_j}|$).  Here the sum $(i,j)$ extends over all independent pairs with separations between $\erre$
and $\erre+d\erre$.  Any weighting which is independent of the local galaxy
density will lead to an unbiased estimate of the correlation 
function, in the sense that the mean value of the estimator, if averaged over 
an ensemble of similar surveys, would approach the true, underlying galaxy 
correlation function.  However, the {\it variance} of individual 
determinations of $\xi$ around this mean {\it will} depend on the particular 
choice of $w_{ij}(r)$; some choices lead to more accurate 
estimates than others.  By assuming that two--point correlations 
dominate over higher orders (essentially the linear regime),
one finds a useful approximation to the {\it minimum 
variance} weights (Saunders et al. 1992; Hamilton 1993; Fisher
et al. 1994a, F94 hereafter), as
\begin{equation}
\label{MVweights}
w_{ij}(\erre) = {1\over 1 + 4\pi \ng \phi(d_i) J_3(\erre)} \cdot {1\over 
1 + 4\pi \ng \phi(d_j) J_3(\erre)}\,\,\,\, ,
\end{equation}
where $\phi(d_i)$ is the survey selection function at the position
of galaxy $i$, and $J_3(\erre) \equiv \int_o^\erre dr \, r^2 
\xi(r)$.  The minimum variance weights themselves, therefore, require
knowledge of $\xi$, i.e. we have a circular argument for which one must
hope that an iterative procedure would be convergent.   We shall instead
adopt an {\it \`a priori} model for $\xi$ and use it to calculate the
weights; given the host of
assumptions employed to arrive at (\ref{MVweights}), this approach should
not be considered as too gross an approximation, as also shown, e.g., by
F94. We model \xr as a power law,
$\xi(r) = \left({5/r}\right)^{1.8}$ 
for $r<30\hmpc$, and assume negligible correlations at larger separations.  
This expression results 
in $\j3(\erre)=15.1\, \erre^{1.2}$, for $\erre\le 30\hmpc$, and
$\j3=894$ at larger separations.  

To test the sensitivity of the results to the adopted model, we also compute
$\j3$ in a different way, i.e. using its definition in terms of the power 
spectrum P(k).  In this case, we have  
$\j3(\erre) \equiv (4\pi r^3/3)\int_0^\infty dk k^2 P(k) W_{th}(k\erre)$, 
where $W_{th}(kr)$ is 
the Fourier transform of the spherical top-hat window function.  We compute 
$\j3$ for the CDM power spectrum normalized to unity in spheres of radius $8 
\hmpc$. [We note that variants of standard CDM, such as an open model 
or a low-density flat model with non--zero $\Lambda$, would provide a more 
realistic power spectrum, but as it is clear from the results of the test, 
this would make no real difference for our purposes].  We find no significant 
deviations between the estimates of \xs computed using the two different 
models for $\j3$, and therefore used the one based on \x in the rest of
the computations.

The weighting function has to be taken into account also when calculating
the mean densities $\ng$ and $\nr$ to be used in eq.~\ref{est1}.  A full 
discussion of the merits of different estimators for the mean density in 
a magnitude-limited sample is presented in paper II.   
In general, an estimate of the mean density, given a weight for each object 
and the survey selection function, would be of the form
\begin{equation}
\ng = \frac{\sum_i w_i} {\int_{V_s} dV\, w(d) \phi(d)} \,\,\,\,	,
\end{equation}
where the sum extends over all galaxies (or random points), while the
integral is extended to the whole survey volume $V_s$.  In practice, 
we are
interested in the ratio of $\nr$ and $\ng$, so that we only need to estimate
for the two sets of points the value of the quantity in the numerator
(i.e. the effective total number of objects).   The same arguments 
leading to eq. (\ref{MVweights}) result in a similar expression 
for the minimum variance, single--point weights: $w_i = 
(1 + \ng \phi(d_i) J_3^{max})^{-1}$ (Davis \& Huchra 1982);
here $J_3^{max}$ is the maximum value of $J_3$.  Note that again we have one
quantity, $\ng$, that ideally should be calculated iteratively.  
However, since in any case our estimator will be unbiased, no matter the choice 
of the weight, we use here - and correspondingly in the pair weighting - the 
mean density as estimated in paper II from this same sample, with the same 
$M_{b_J}\le-12.4
+5\,{\rm log\, h}$ cut. 

In our specific case, there is an additional complication arising from 
the varying sampling of the individual 
Optopus fields, as discussed in \S~\ref{sec-data}.  
This requires the introduction of another weight, $W^i$, 
to correct for the missed 
galaxies in each field.  Following the same reasoning as detailed 
in V98, we define the completeness in each field $C^i$ as 
\begin{equation}
C^i = {{N^i_{Z}}\over{(1-f_*)\, N^i_T}},
\label{Cf}
\end{equation}
where $N^i_Z$ and $N^i_T$ are the number of secured galaxy redshifts and the number of objects classified as galaxies in the parent photometric catalogue,
respectively, in each field $i$, while $f_*=0.122$ is the observed fraction of misclassified stars. 
We therefore multiply the $w_i$ corresponding to each galaxy by the weight
pertaining to its field, that will be given by $W^i=1/C^i$ .   
This is not necessary for the random sample, that, by construction, 
is uniform over the Optopus fields.

Eq. (\ref{est1}), is based on the definition of 
$\xi$ (e.g. Peebles 1980), in which it is understood that $\ng$ 
represents the universal mean galaxy density.  Any sample 
mean density will vary about this value with a dispersion 
at least as large as expected from simple Poisson statistics 
and, in fact, clustering on the scale of the survey volume
will increase this dispersion.  The precision of the 
estimate of $\xi$ at small values (large scales) is limited 
by this uncertainty, $\delta\ng$ (the number of points in the 
random catalog $N_r=V\cdot \nr$ may always be sufficiently increased to eliminate 
its density as a source of uncertainty, and this is in fact what we do 
here by using typically $N_r=100,000$).  Given that $\ng$ 
appears linearly in estimator (\ref{est1}), one would
conclude that the precision on $\xi$ is also linear
in $\delta\ng$.  In fact, this is {\it not} correct, 
as emphasized by Hamilton (1993) and Landy \& Szalay (1993), who suggested 
alternative estimators in which the precision goes as $\delta\ng^2$,
as would seem more appropriate for the second moment of a distribution. 
\begin{figure*}
\resizebox{\hsize}{!}{\includegraphics{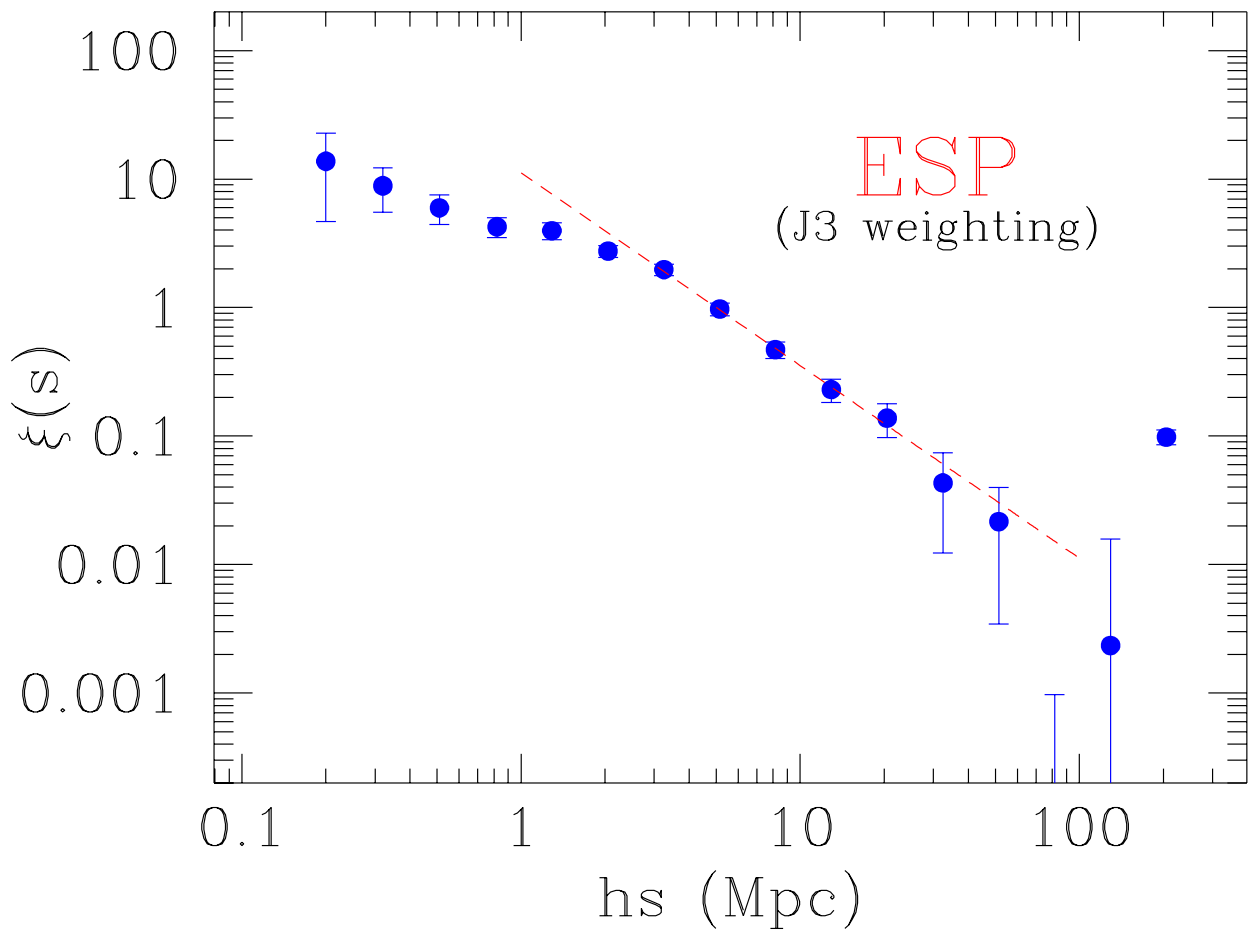}}
\caption{Redshift--space correlation function from the whole ESP
magnitude--limited catalogue, computed using the 
minimum--variance, $J_3$--weighting scheme.  For reference, the dashed 
line gives an arbitrary power-law function with 
$r_o=5\hmpc$ and $\gamma=1.5$.}
\label{xis}
\end{figure*}
We have computed \xis using (\ref{est1}), and with Hamilton (1993)
estimator, and found for our survey equivalent results within the errors.  
This is partly in contrast to what was found by Loveday et al. on the 
Stromlo--APM survey (1995), but agrees to the results of Tucker et al. 
(1997) on the larger LCRS (cf. their Figure 1). The results we shall show 
in the following are therefore all based on the Davis \& Peebles
(1983) estimator.

\subsection{Error estimation and model fitting}

Statistical error bars for our estimates of two-point correlation functions 
are computed using bootstrap resampling as discussed by Ling et al. (1986).
F94 have discussed in detail the reliability of the
bootstrap technique in estimating errors for correlation functions; they
show that, typically, bootstrap errors are a good description of true rms 
errors for separations smaller than $\sim 10\hmpc$, while on larger scales 
they tend to overestimate them by a factor of about 1.5.  In the same
work, they carefully discuss the technique for  fitting of models to
correlation functions,
where the data points are not statistically independent.   
We use the same procedure here, which we briefly summarize in the
following. Essentially, we create $N_B=100$ "perturbed" realizations 
of the sample, 
by randomly sampling the original catalog {\it with replacement}, (the "bootstrapping").  
By subjecting each of these realizations to the same analysis, we obtain a set 
of $N_B$ correlation functions: $\xi_k (r)$, with 
$k =1, N_B$.\footnote{Note that here we shall fit 
models to one--dimensional quantities only -- i.e. \xis or projections 
of \xip} We refer to the set of 
separations $r_i$, one for each bin of size $\Delta r$,
as {\it separation space}.  The correlation function $\xi(r)$
is a vector, $\vec{\xi^S}$ in this space with 
components ${\xi(r_i)}$.  A covariance matrix in separation space may  
be constructed as
\begin{equation}
\Cmat(r_i,r_j) = <(\xi(r_i)-{\bar\xi(r_i)})(\xi(r_j)-{\bar\xi(r_j)})>_{realiz}, 
\end{equation}
where '$<>_{realiz}$' indicates an average over bootstrap
realizations and ${\bar\xi(r_k)}$ is the ensemble average of the 
correlation function at separation $r_k$.  Since the values of $\xi$ at 
two different separations are correlated, $\Cmat$ is non--diagonal.  
For this reason one cannot do a straightforward $\chi ^2$ fit of 
a model to the observed points. 
However, $\Cmat$ is symmetric ($\Cmat=\tilde \Cmat$) and real, and 
therefore hermitian, i.e.
such that it can be diagonalized by a unitary transformation if its
determinant is non--vanishing.  What happens
in practice, is that the estimated functions are oversampled, so 
that the effective number of degrees of freedom in the data is smaller 
than the number of components of $\vec{\xi^S}$.  This makes $\Cmat$ singular,
and one has to apply singular value decomposition (as in F94),
to recover its eigenvectors.   These define the required transformation
$T$, to the orthogonal space, i.e. 
$\Cmat \rightarrow \Dmat\ = \tilde \Tmat\cdot \Cmat\cdot \Tmat$ and 
$\vec{\xi^S} \rightarrow \vec{\xi^D} = \Tmat\cdot\vec{\xi^S}$,
where now $\Dmat$ is diagonal and the components of $\vec{\xi^D}$ 
are independent.
In this space we can therefore define in the usual way the quantity 
$\chi^2$ with respect to our model parameters, and minimize it to find 
their best--fit values.  

\section{The Redshift--Space Correlation Function $\xi(s)$}
\label{sec-xis}

\subsection{Optimal weighting} 

\begin{figure*}
\resizebox{\hsize}{!}{\includegraphics{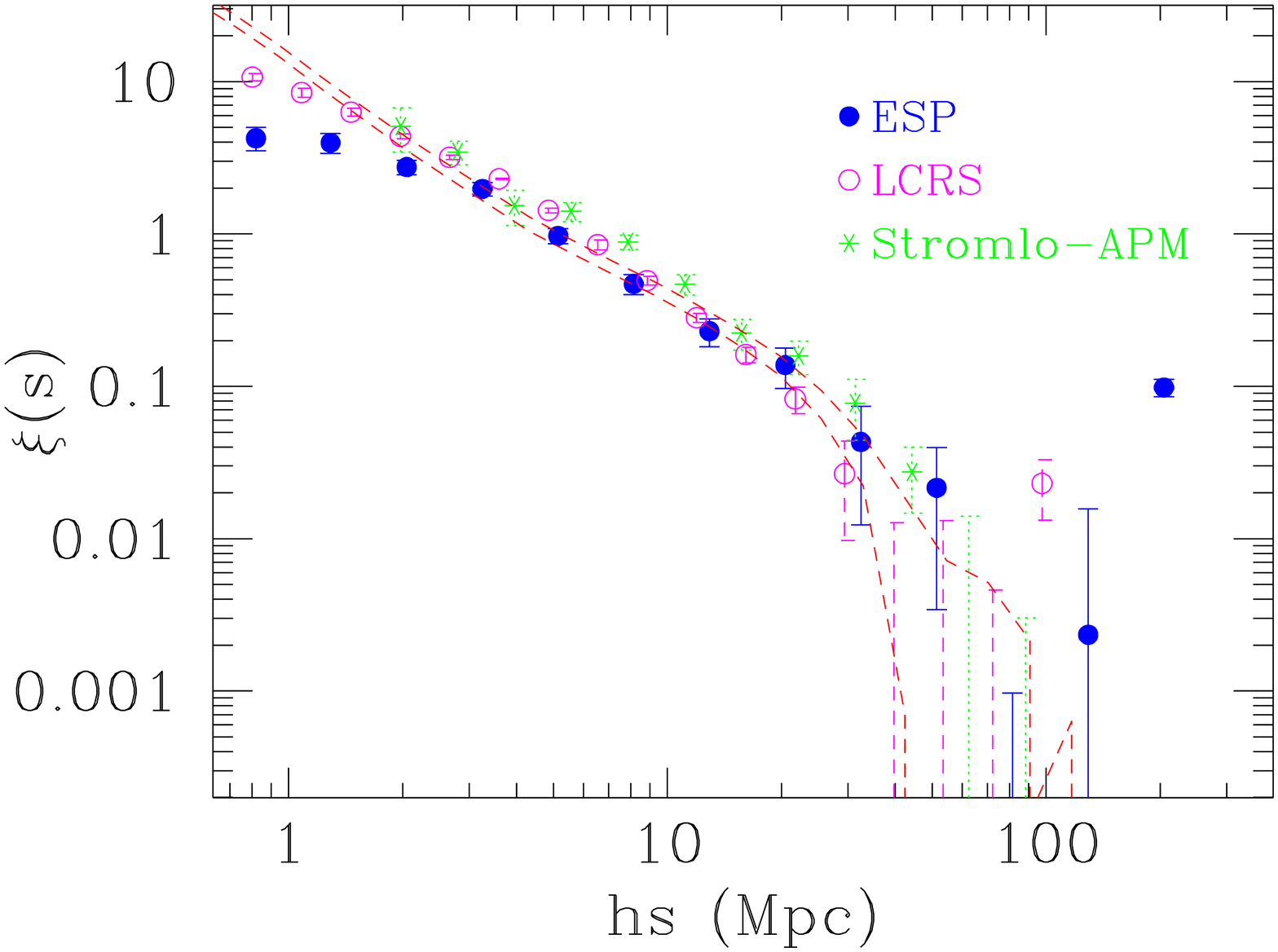}}
\caption{\xis on large scales from the ESP, LCRS (Tucker et al. 1997), 
and Stromlo--APM (Loveday et al. 1992b), redshift surveys.  
The dashed lines show the real--space \xir as obtained
through deprojection of the angular correlation function for the APM
galaxy catalogue, for two clustering evolution models (Baugh 1996).}
\label{xis-surv}
\end{figure*}

To maximise the large--scale signal from the data,we first estimate
\xis from the whole ESP redshift catalogue,
preparred and trimmed as described in \S~\ref{sec-data}.
Figure~\ref{xis} shows the result of applying the $J_3$ optimal--weighting
estimator to this sample, containing 2850 galaxies.
Between 3 and $50\hmpc$ \xis is well described by a shallow 
($\gamma \simeq 1.5$) power law, with a redshift--space correlation length 
$s_o\simeq 5\hmpc$.   On larger scales, it smoothly breaks down,
crossing the zero value between 60 and $80\hmpc$.   The bin centered
at $80\hmpc$ has indeed a negative value, although it is
less than 1$\sigma$ below zero.  On larger scales ($r>100\hmpc$), 
there is marginal evidence that the amplitude of \xis rises up again, 
a behaviour that seems to be shared also by other clustering data, as 
we show below.   On scales smaller than $3 \hmpc$, \xis flattens significantly,
so that a single power--law is certainly not a fair description over 
the whole explored range.

A comprehensive comparison of the best-fit results of a power--law
\xis to redshift--space correlation functions from several surveys
is given in Willmer et al. (1998).  As one could easily guess from our
Figure~\ref{xis}, these results are strongly dependent
on the range of scales over which the fit is performed, that often is
not the same for the different data sets.  This makes the comparison of
the crude best-fit values rather inconclusive.  Equivalently difficult
is any physical interpretation of the measured amplitudes and slopes,
given that the fit sometimes includes scales where power suppression 
by virialised structures (``Fingers of God'') is very effective 
($s<2\hmpc$), while in other cases it starts above these,
putting more weight on less nonlinear scales.  Given these ambiguities,
here we do not perform a formal fit to the ESP \xis.  We shall compute 
a proper fit only to the real-space correlation function, for which 
between 0.4 and $10 \hmpc$ a power-law shape is a rather good 
description of the data, and where the observed shape
is at least freed of one major distorting effect.

In Figure~\ref{xis-surv}, we compare the $J_3$ estimate of \xis from
the ESP on large scales to those computed with the same technique from the
LCRS and the Stromlo--APM surveys\footnote{We limit our explicit comparison 
to these two surveys, as the LCRS is the only other redshift survey
with depth and angular coverage comparable or superior to the ESP.
The Stromlo--APM survey, on the other hand, is less 
deep by $\sim 2$ magnitudes and is sparsely sampled, but  
represents an interesting comparison because of its large solid angle and
the fact of being selected from the same photographic material 
(IIIaJ plates).}.  The agreement between the three estimates
is very good between 2 and $30\hmpc$, given also the different 
galaxy selection functions of the corresponding surveys.  On larger
scales we can note that the ESP and Stromlo--APM
\xis seem to show slightly more power
than the LCRS. This evidence for
low--amplitude power on large scales, with a smooth decay
from the power--law shape, resembles that observed
in the IRAS 1.2Jy redshift survey (F94).
One could probably explain the differences in Figure~\ref{xis-surv} 
with the different selection criteria in the three redshift catalogues:
the LCRS is selected in $r$ (with the additional complication of a
surface--brightness cut, whose effect is not fully clear), and thus should
tend to favour earlier types on the average, that we know reside 
preferentially in high--density regions.  On the other hand, 
the $b_J$ band used for the ESP (so as the IRAS infrared band),
is expected to better select star--forming objects in low--density 
regions.  It should probably be expected that ESP galaxies, as well as
IRAS galaxies, are better tracers of very large--scale, low--amplitude 
fluctuations.

In Figure~\ref{xis-surv}, we also plot (dashed lines), the real--space
correlation function obtained by deprojecting the angular correlation
function $w(\theta)$ from the APM galaxy catalogue (Baugh 1996).   This is 
computed in the case of clustering fixed in comoving coordinates (bottom
curve), or growing according to linear theory (top).  
While the details of the real--space correlation function will be analysed 
in the following section directly from the ESP itself, here we have the
chance to make already a few interesting remarks.

\begin{figure*}
\resizebox{\hsize}{!}{\includegraphics{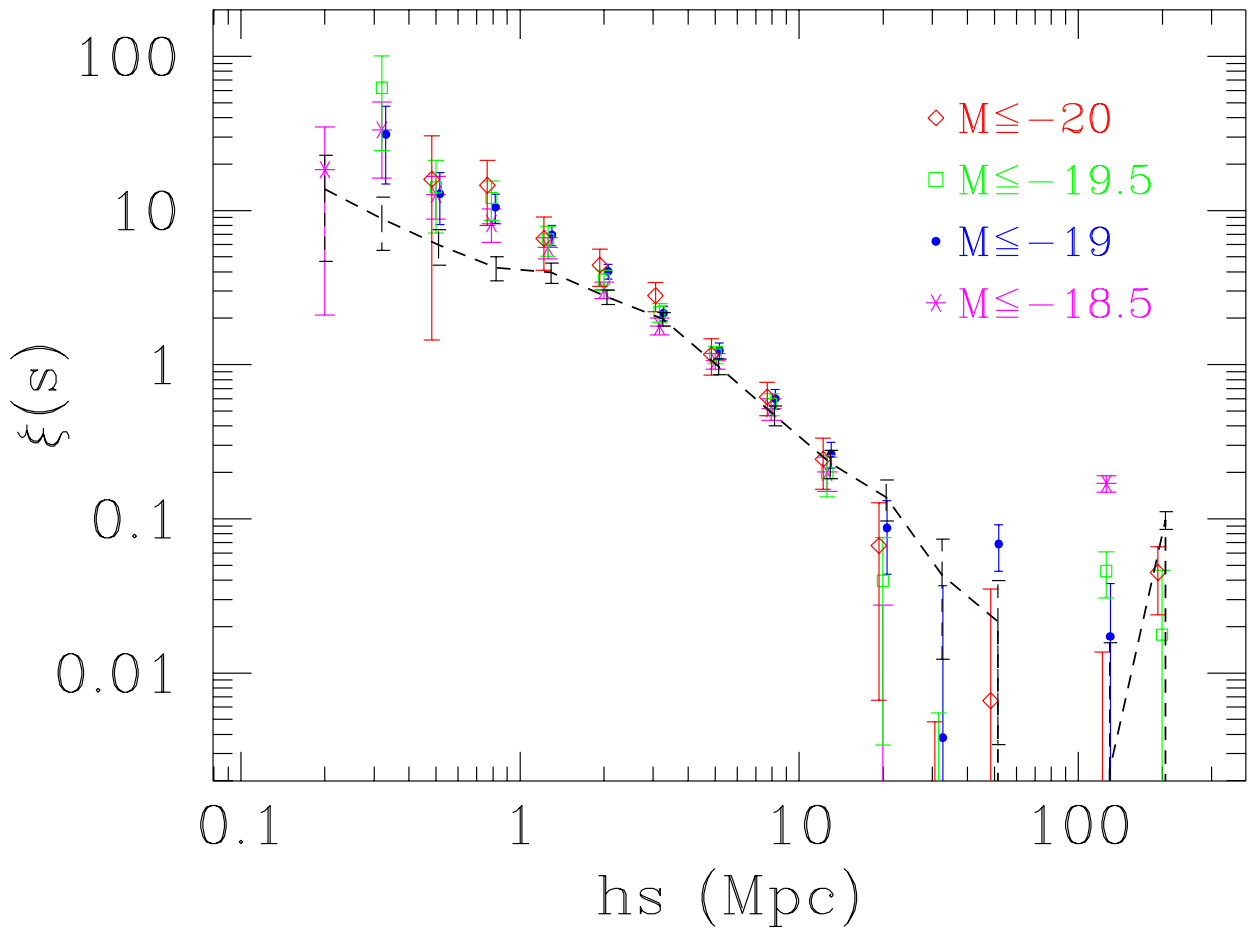}}
\caption{Redshift--space correlation function from four 
representative volume--limited
samples extracted from the ESP as described in Table 1.
The dashed line with error bars reproduces the $J_3$-weighted estimate
from the whole survey.  For clarity, points for $M\le-19$ and $M\le-20$
are displayed with a shift in log(s).
}
\label{xis_vlim}
\end{figure*}

First of all, the scale of the break in both \xis and \xir is consistently
indicated by the different surveys to be between 50 and $90\hmpc$,
with the ESP clearly pointing to a value $>50\hmpc$.  The agreement between
these new redshift data and the APM \xir, directly implies that the
large--scale power originally seen in the angular correlation
function $w(\theta)$ from the APM (Maddox 
et al. 1990) and EDSGC (Collins et al. 1992), galaxy catalogues 
is not significantly enhanced by errors in 
the magnitude scale, as claimed by some authors (e.g. Bertin \&
Dennefeld 1997).  In fact, Figure~\ref{xis-surv} demonstrates that similar
power is seen in redshift
survey data, either based on the same $b_J$ plates (ESP) or CCD
photometry (LCRS).  Also concerning the detailed {\sl shape} of 
galaxy correlations above
$3 \hmpc$, independent data sets show a high degree of unanimity: 
if one ideally
extrapolates to larger scales the `classic' $\sim -1.8$ slope
observed in real space below $3\hmpc$ from the APM \xir (but also
from previous angular projections, see e.g. Davis \& Peebles 1983),
all surveys are systematically above this extrapolation.  To reproduce
this feature (described in the literature as a `shoulder' or a
`bump', see Guzzo et al. 1991, and Peacock 1997), a rather steep 
($\sim k^{-2}$) power spectrum P(k) is required
(Branchini et al. 1994, Peacock 1997).

A second important point about Figure~\ref{xis-surv} concerns the
effect of redshift--space distortions.  One can see how small
is the amplitude difference between the redshift--space correlation
functions and the APM real--space \xir on large scales.  
A linear amplification is indeed expected as a result of
coherent flows towards large--scale structures.  This can
be approximately expressed
as $\sim$ $1 + { 1 \over 2 } {\beta} + { 1 \over 5 }
{\beta^2}$, where $\beta=\Omega_o^{0.6}/b$ and $b$ is the linear
biasing factor (Kaiser 1987).  Figure~\ref{xis-surv} implies, 
within the errors, a value for this factor very close to 1.
For example, a value of 1.2, consistent with the observed data, 
would yield $\beta=0.35$.

It is interesting to note that the Stromlo--APM \xis is the only
data set that around $\sim 10\hmpc$ seems to show a significant
amplification with respect to \xir.  (Note that this comparison
is actually the safest, being the APM catalogue the parent photometric
list of the Stromlo--APM redshift survey).  
The interpretation of this effect is however complicated by
what we have pointed out in 
Paper II, where we showed how the mean density in this redshift survey
is about a factor of 2 lower with respect to that deduced from deeper samples.
The cause for this seems to be a negative density fluctuation that we 
clearly detect in the ESP data below $z<0.05$, and that is also visible in 
other surveys.  This "local" underdensity, 
is shown to be the explanation for both the low normalization of the 
Stromlo--APM
luminosity function (Loveday et al. 1992a), and the steep number counts
observed at bright magnitudes in several bands (e.g. Maddox et al. 1990).
It is not trivial to understand how this systematic 50\% underestimate of the 
mean density would affect the two--point correlation function, although it 
clearly reduces by the same amount the number of pairs expected at a given 
separation.  If this deficit is evenly spread on all scales, then the measured
\xis will be practically unaffected, however if it corresponds, for example,
to a region with larger-than-average voids, the net effect will be to boost
up \xis by some amount.  This would be sufficient to produce the observed 
discrepancy between \xis and \xir, and further demonstrates how extended
and accurate the data have to be in order to extract dynamical information 
as the value of $\beta$.

\subsection{Volume--limited estimates}

While the $J_3$--weighted estimate has the advantage of maximising the
information extracted from the available data, it has some important
drawbacks that have been not always appreciated in previous applications.
The main problem is that by its own definition it inevitably mixes
the contribution of galaxies with different luminosities.  This would
not be a problem if galaxy clustering were completely independent of
luminosity, i.e. if each galaxy traced fluctuations independently
from its own absolute magnitude.   As we discussed in \S\ref{sec-est},
this technique weighs pairs based both on their separation and on their
distance from the observer.  This implies that the main contribution
to small--scale correlations comes from low--luminosity pairs, that
are numerous (and dense) in the nearby part of the sample and thus
better trace clustering at small separations.   On the contrary, the
estimate of \xis on large scales is dominated by the contribution from
luminous objects, that are in fact the only ones detectable in the
distant part of the survey.   Clearly, if there is any luminosity
dependence of clustering, this technique will tend to modify in
some way the shape of \xis.   In particular, if luminous galaxies
are more clustered than faint ones, the $J_3$ weighting will tend to
give a shallower slope for a power--law shaped correlation
function.  Therefore, while this method is certainly
optimal for maximising the clustering signal on large scales (and
so our conclusions on the large--scale shape of \xis from the previous
section should not be affected), it might be dangerous to draw from its
results far--reaching conclusions on the global shape of \xis,
especially on small scales.   A wise way to counter--check the results
obtained from the optimal weighting technique, is that of estimating
\xis also from volume--limited subsamples extracted from the survey.
In this case, each sample includes a narrower range of luminosities
and no weighting is required, the density of objects being the same
everywhere.  This means that in our case $w_i(\erre)\equiv 1$, and
only the $W^i$'s have to be taken into account.

\begf
\resizebox{\hsize}{!}{\includegraphics{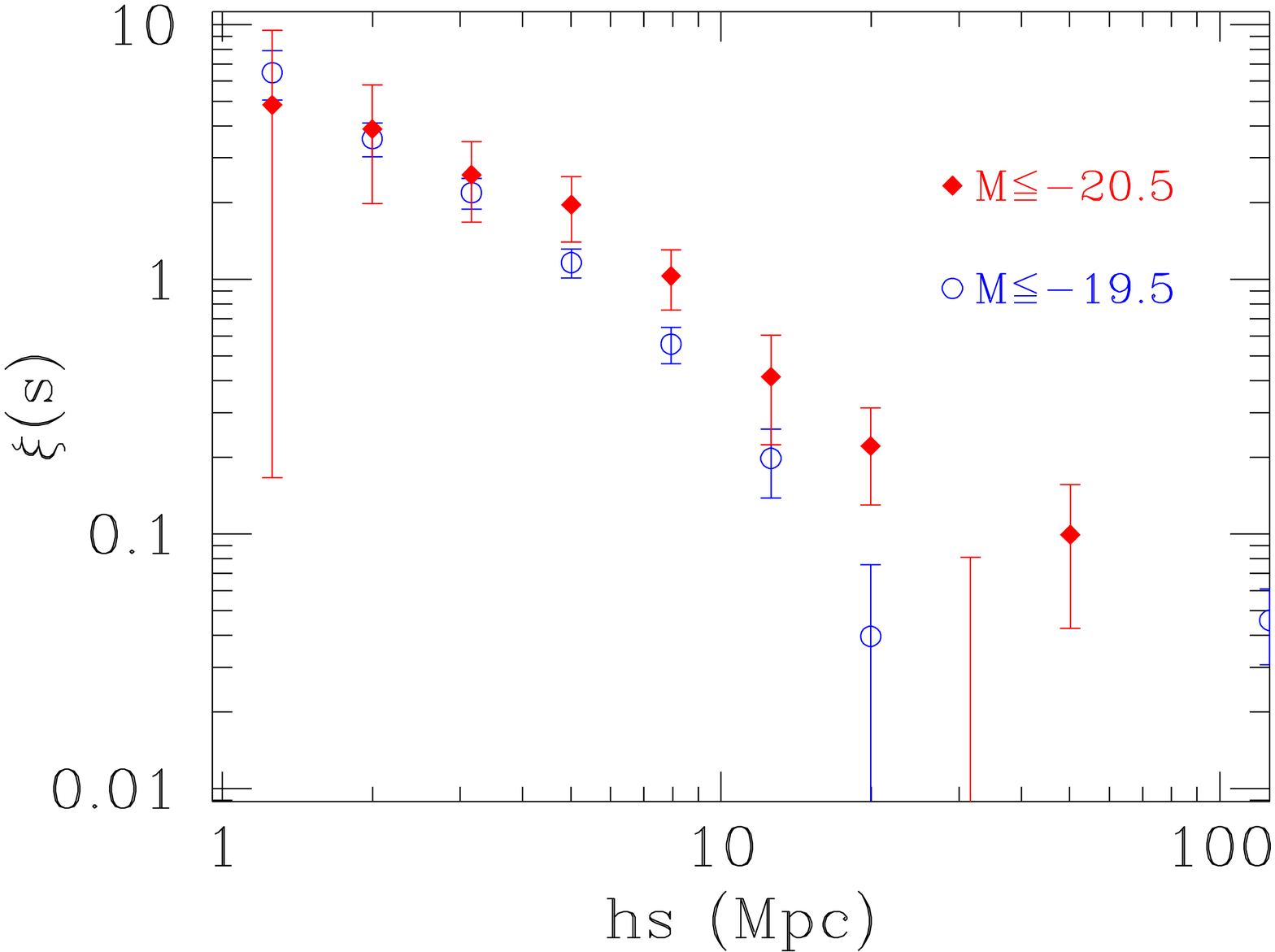}}
\caption{A further test for luminosity dependence of clustering in 
redshift space. \xis for the brightest 
($M\le-20.5$) volume-limited subsample of the ESP is
compared to one of the main samples from the previous figure.
Note that the scale has been changed to better evidence  
the difference in amplitude on intermediate scales.
}
\label{xis_vlim_195_205}
\endf

We have performed this exercise on the ESP, and the 
results for the four main samples defined in Table 1
are displayed in Fig.~\ref{xis_vlim}.
In the figure, we also reproduce the $J_3$--weighted estimate (dashed line).
To ease visualization, 
the data points for the $M\le-20$ and $M\le-19$ samples have been
shifted by a constant value (negative and positive respectively) in log$(s)$.
It is clear how for the ESP data the $J_3$--based method produces
a \xis which is shallower below $3\hmpc$.  On larger scales the
shape is consistent between the two methods, and for $s>20\hmpc$
the optimal weighting performs better in terms of signal--to--noise,
than the single volume--limited estimates.  The observed \xis is
in practice the result of the convolution of the real--space
correlation function \xir with the distribution function of
pairwise velocities.  This means that the small--scale flatter 
shape of \xis in the case of the magnitude--limited sample could be
in principle produced either by a smaller \xir or by a higher pairwise 
velocity dispersions.  We find (Guzzo et al., in preparation), that 
most of the effect is produced by this second factor: the small--scale
pairwise velocity dispersion between 0 and $1 \hmpc$, is higher when 
measured on the whole survey using the $J_3$--weighted method. 
We measure $\sg \simeq 600 \kms$ using
the $\j3$ estimate from the whole sample, while the
volume--limited samples give values between 300 and $ 400
\kms$, depending on the infall model adopted.

From this plot alone there is no evident sign of a luminosity dependence
of \xis, since the four estimates are virtually the same within the errors.
From their analysis of the SSRS2 sample, Benoist et al. (1996), also find 
that there are negligible signs of luminosity segregation for 
galaxies fainter than $M^*$, which in our case corresponds to $-19.6$.
They also show, however, 
that a clear effect becomes visible for absolute magnitudes brighter than 
$\sim -20$.  To check for this effect in more detail, in 
Fig.~\ref{xis_vlim_195_205} we plot 
\xis for the brightest sample that can be selected from the
ESP while keeping a reasonable statistics, with $M\le-20.5$.  This estimate
is compared to that for the $M\le -19.5 \simeq M^*$ sample from the
previous figure.  Here, we see a significant increase in the amplitude 
of \xis for the more luminous galaxies, for separations above $3-4\hmpc$.
Even if we do not perform a formal fit for the reasons discussed
above, we see that now \xis passes through unity around $r\sim 8\hmpc$,
i.e. has an amplitude very consistent to that observed 
by Benoist et al. (1996) for galaxies with similar luminosity in the SSRS2.

These authors also point out how the detection of luminosity segregation 
is complicated by volume effects (different samples covering different 
volumes), and by redshift--space distortions.  For these reasons, the
results of Figures~\ref{xis_vlim} and \ref{xis_vlim_195_205} are 
probably meaningful for what concerns the shape and amplitude of \xis on 
scales $>4-5\hmpc$ (given also the negligible amount of redshift--space
amplification that we have shown in Figure~\ref{xis-surv}).  On smaller
scales, however, clustering and dynamics mix up in a non--trivial way
for the different samples, so that the simple redshift--space correlation 
function is possibly hiding a large amount of information.   Guzzo et al. 
(1997) have shown how to disentangle some of these effects by studying the 
differences of clustering in real space in the case of the Persus--Pisces
survey.  More recently, Willmer et al. (1998) have studied \xir and \xis
in the SSRS2 survey for different luminosity subsamples.
We shall see in \S~\ref{sec-xir} how also in the case of the ESP,
when studying \xip and its real--space projection \wp, it is possible 
to evidence some weak positive trend of clustering with luminosity 
also at small separations.

\section{$\xi(r_p,\pi)$ and the Real--Space Correlation Function}
\label{sec-xir} 

So far, we have considered the correlation among galaxies as located in
{\it redshift space}, where a map of the galaxy distribution is distorted 
due to the effect of peculiar velocities, whose contribution adds to
the Universal expansion to produce the redshift we actually measure.
The effect of redshift--space distortions can be shown explicitly through 
the correlation function \xip, where the separation vector of a pair of objects
is split into two components: $\pi$ and $r_p$, respectively parallel and
perpendicular to the line of sight.  Given two objects with observed radial 
velocities $v_1=cz_1$ and $v_2=cz_2$, following  F94 we 
define the line of sight vector 
${\bf l} \equiv (\vva + \vvb)/2$ and the redshift difference
vector ${\bf s} \equiv \vva - \vvb$, leading to the definitions
\begin{equation}
\pi \equiv {{{\bf s} \cdot {\bf l}}\over{{\rm H_0} |{\bf l}|}}\quad\quad
r_p^2 \equiv {{\bf s} \cdot {\bf s}\over {\rm H_0^2}} - \pi^2\, .
\label{f94-def}
\end{equation}
We can then generalize estimator
(\ref{est1}) to the case of \xip, now counting the number of pairs in a 
grid of bins $\Delta r_p$, $\Delta\pi$.   

\begf
\resizebox{\hsize}{!}{\includegraphics{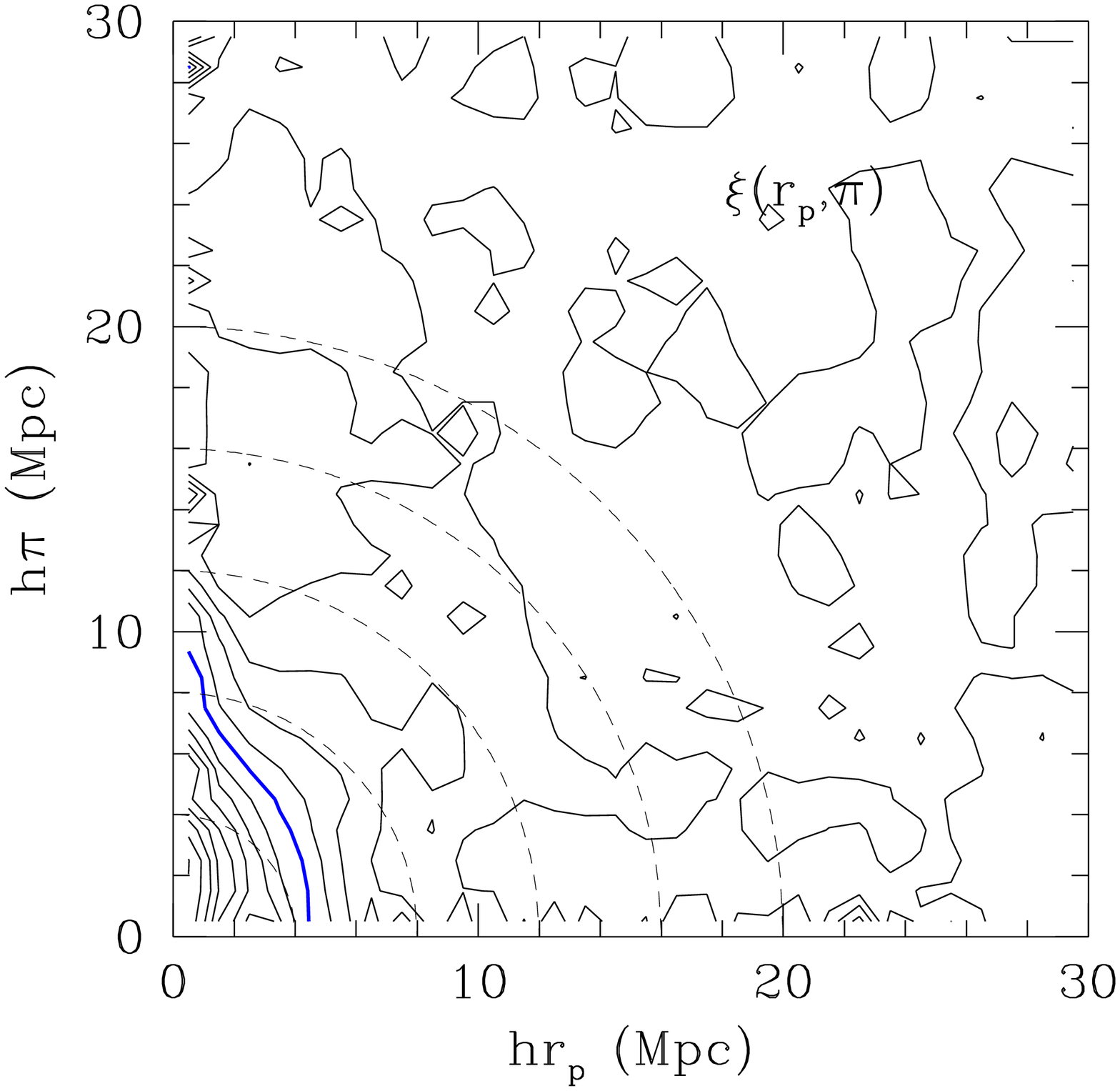}}
\caption{\xip for the ESP magnitude--limited sample.  
The heavy contour corresponds to $\xi=1$; 
for higher values of $\xi$, contours are logarithmically spaced, 
with $\Delta \log_{10} \xi = 0.1$; below $\xi=1$, they are linearly spaced 
with $\Delta \xi=0.2$ down to $\xi=0$.  The dashed contours represent
the isotropic correlations expected in the absence of peculiar
velocities.  The map has been Gaussian-smoothed with an
isotropic filter of width 3~$\hmpc$.}
\label{csipz_j3}
\endf
\begf
\resizebox{\hsize}{!}{\includegraphics{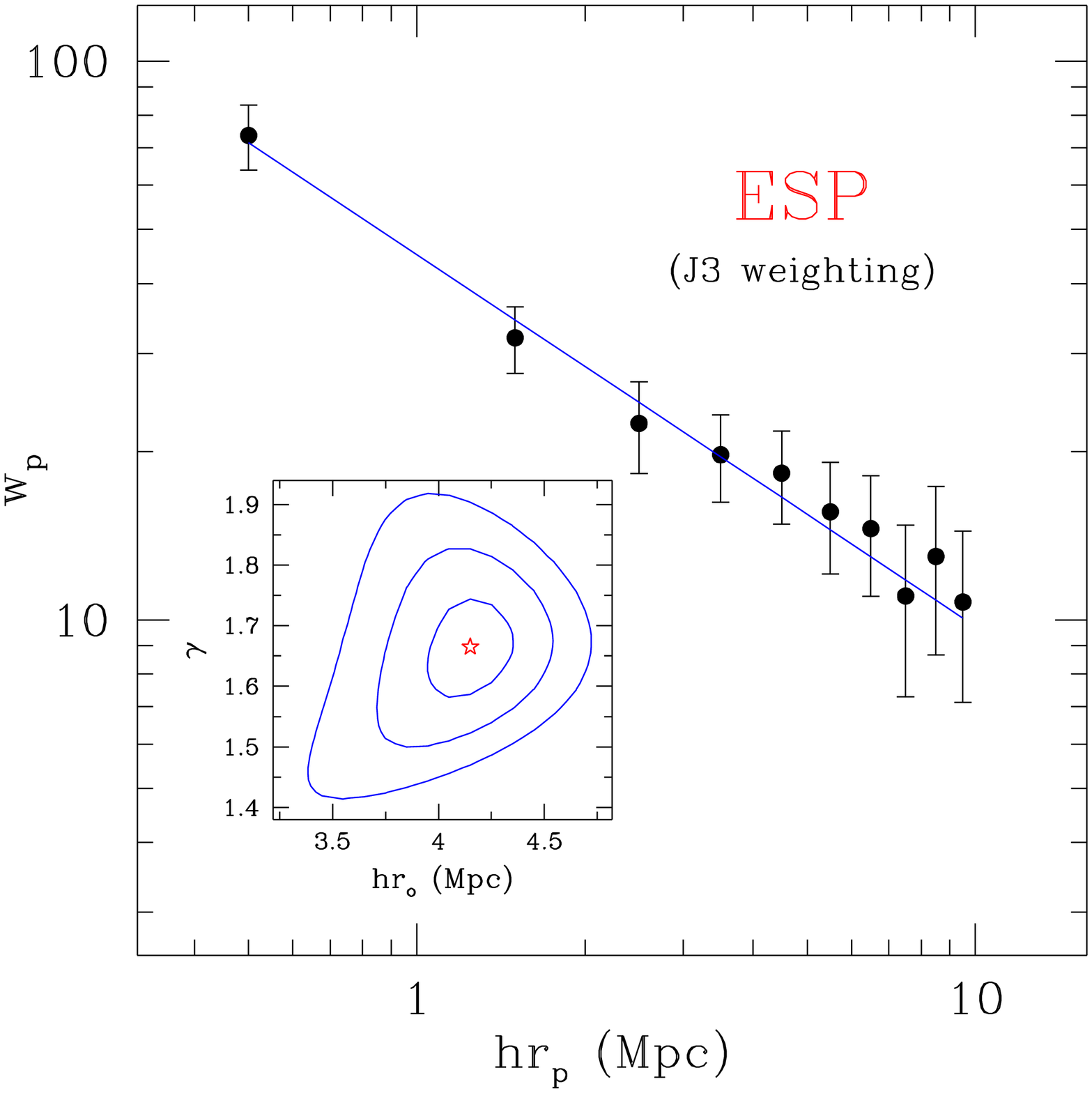}}
\caption{Results of the fit of a power--law \x to the projected function
\wp (filled circles), for the whole magnitude-limited redshift sample.
The inset shows the corresponding best-fit values for the correlation length
$r_o$ and the slope $\gamma$, together with the single-parameter confidence
contours at the 1-, 2- and 3-$\sigma$ levels.}
\label{wp_j3}
\endf
Figure~\ref{csipz_j3} displays the observed \xip\ for the ESP survey, 
estimated from the magnitude-limited sample using the same minumum 
weighting technique used for computing \xs, but binned linearly into
bins of $1\hmpc$.  As expected, the iso--correlation
contours of \xip are stretched along the line of sight ($\pi$) at small
separations, with respect to the isotropic case (dashed circular contours).
This is the effect of the relative velocity dispersion of galaxy pairs,
$\sigma_{12}$,
and includes the contribution of both the large velocity dispersion cluster
galaxies, and the cooler field population.   A careful estimate of 
$\sigma_{12}$ 
and a discussion of the small-scale dynamics in the ESP survey will be 
presented in a separate paper.
Here we limit the
analysis of \xip to the task of recovering the real-space correlation 
function \xr.

The standard way to recover \xr is to project \xip\ onto the $r_p$ axis, in 
this way integrating out the dilution produced by the redshift-space
distortion field.  The resulting quantity is
\begin{equation}
w_p(r_p) \equiv 2 \int_0^{\infty} dy\, \xi(r_p,\pi) = 
2 \int_0^{\infty} dy \, \xi\left[(r_p^2 + y^2)^{1/2}\right]\, ,
\label{wp}
\end{equation}
where the second equality follows from the independence of the
integral on the redshift-space distortions.  In the
right-hand side of the equation, $\xi$ is the {\it real--space\/} 
correlation function, evaluated at $r=(r_p^2 + y^2)^{1/2}$. 
With a power-law model for $\xi(r)$, i.e. $\xi(r) = (r/r_0)^{-\gamma}$ 
the integral can be computed analytically, yielding
\begin{equation}
w_p(r_p)=r_p\left({r_0\over r_p}\right)^\gamma {\Gamma({1\over 2})\,
\Gamma({\gamma-1\over 2}) \over \Gamma({\gamma\over 2})}
\end{equation}
where $\Gamma$ is the Gamma function.   
The upper integration limit in Eq.~(\ref{wp}), $\pi_{up}=\infty $, is chosen in 
practice so as to be large enough to produce a stable estimate of
$w_p$.   As in previous works (Guzzo et al. 1997), we find \wp\ to be quite 
insensitive to $\pi_{up}$ in the range $ 20\hmpc < \pi_{up} < 25 \hmpc$ 
for $r_p < 10 \hmpc$.  In fact, for small values of $r_p$, it is important
not to choose a too small upper limit, since this would leave out small-scale
power that is present at large $\pi$'s due to the stretching.
On the other hand, the choice of too large a value for $\pi_{up}$, has the
only effect of adding noise into $w_p$.

\begin{figure*}
\resizebox{\hsize}{!}{\includegraphics{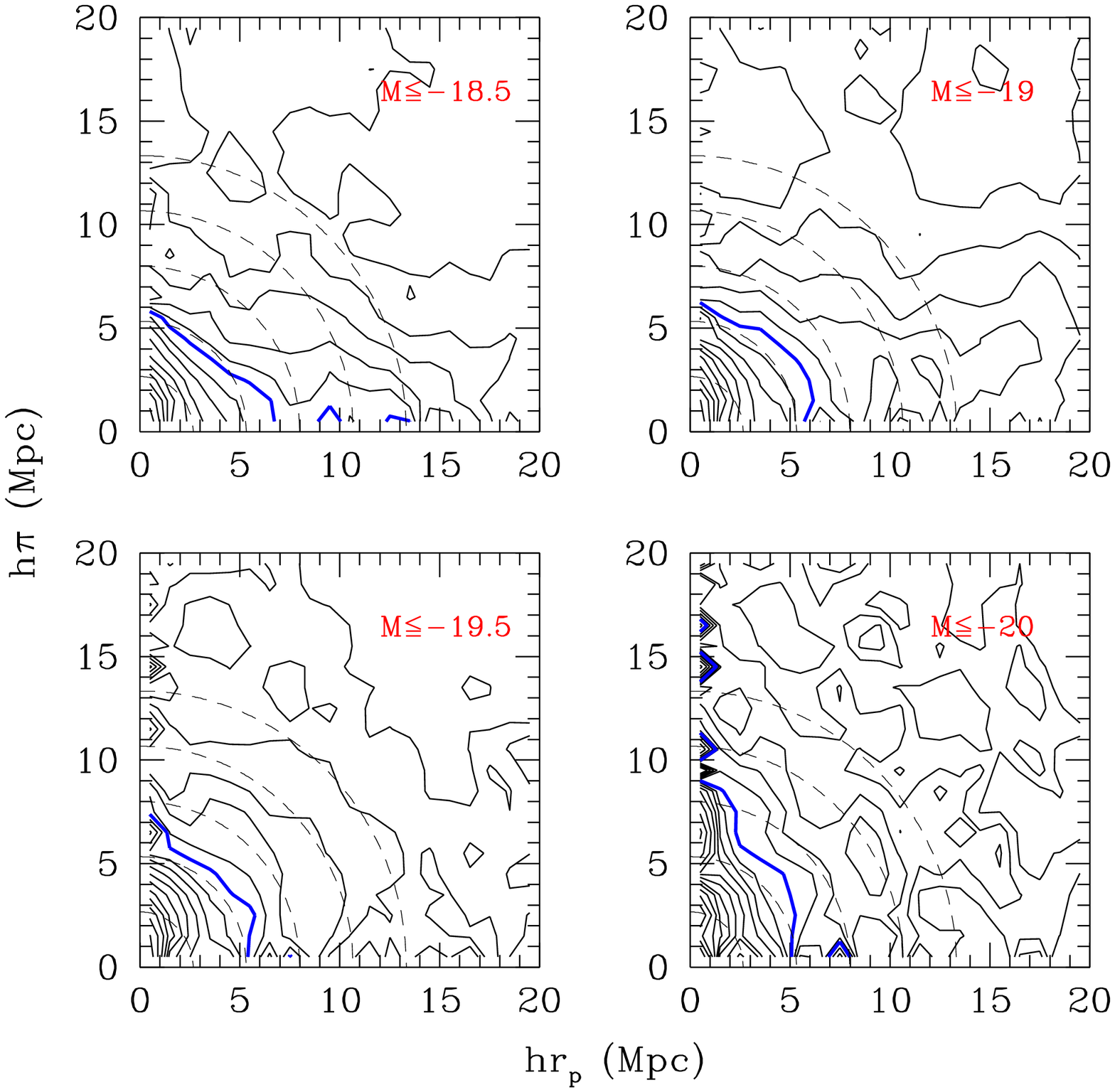}}
\caption{\xip for the four main volume--limited subsamples of the 
ESP catalogue.  Contour levels are defined as in Figure~\ref{csipz_j3}.}
\label{csipz_vlim}
\end{figure*}
\begin{figure*}
\resizebox{\hsize}{!}{\includegraphics{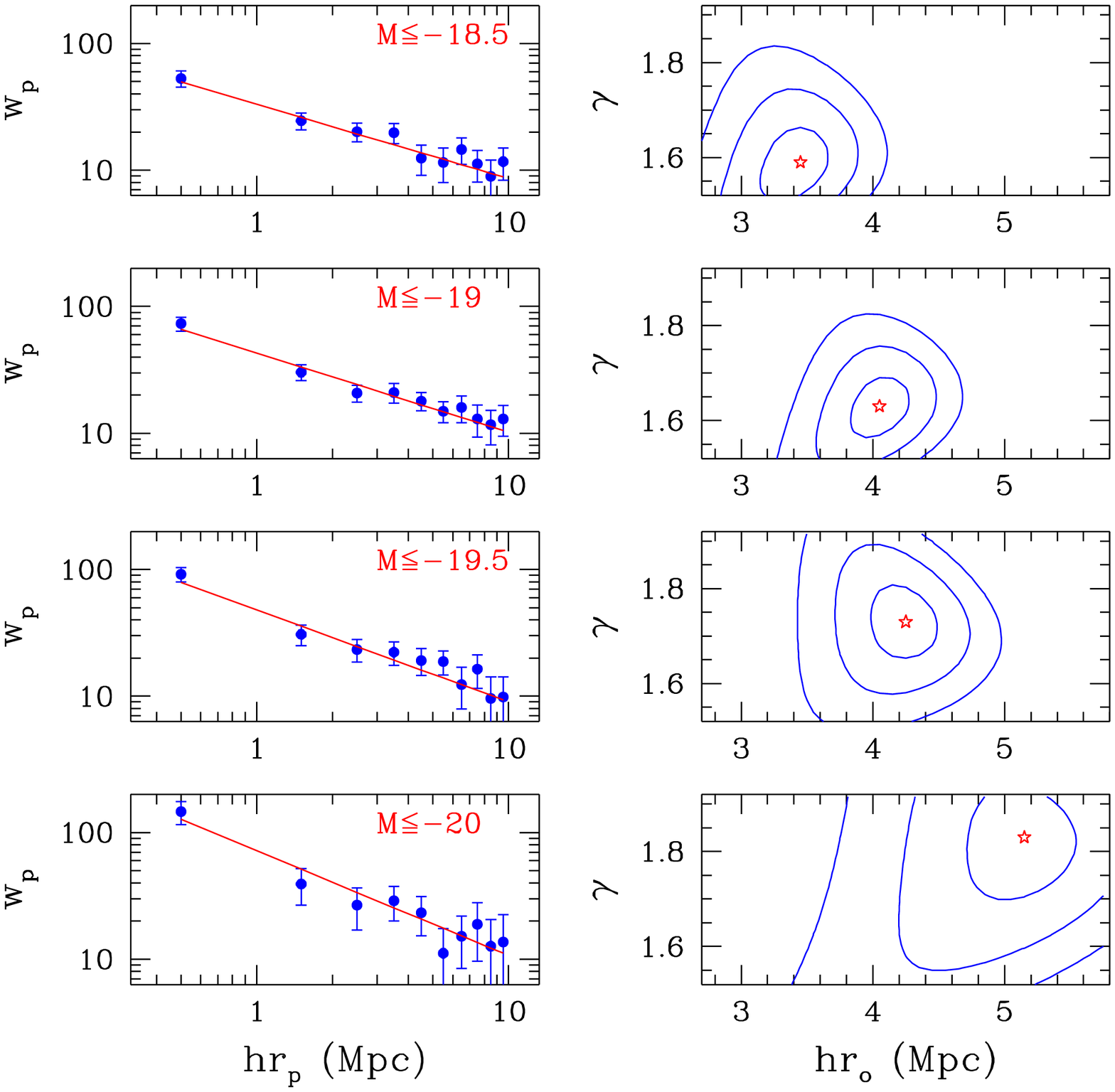}}
\caption{Best--fit results of a power--law \xir to the projected
function \wp for the same four volume--limited samples of the previous
figure.  Left panels give the observed \wp with the corresponding best-fit
power law.  Right panels show the best-fit $r_o$ and $\gamma$ values, together
with their 1-, 2-, and 3-$\sigma$ confidence levels.}
\label{wpfit_vlim}
\end{figure*}

The computed \wp\ for the whole magnitude--limited ESP catalogue is shown 
in Figure~\ref{wp_j3} (solid circles), together with the results of the fit of a power-law model for \xr (solid line and inset).  
Error bars are as usual computed over 100 bootstrap realizations of 
the sample.  The corresponding best-fit values
are $r_o=4.15^{+0.20}_{-0.21} h^{-1}\,$ Mpc and 
$\gamma=1.67^{+0.07}_{-0.09}$.   These values are smaller than those 
measured from the LCRS, for which Jing \& B\"orner (1996) obtain $r_o=5.06 \pm 
0.12$ and $\gamma=1.862 \pm 0.034$.  The reason for this difference is 
probably to be found in the different galaxy populations that these two 
surveys preferentially select, an effect possibly enhanced by the
faint limiting magnitude of the two surveys.  In fact, from the blue-selected
SSRS2 (limited to $m=15.5$, i.e. 4 magnitudes brighter than the ESP), Willmer
et al. (1998) measure a real--space correlation function with $r_o=5.36 \pm 
0.33$ and $\gamma=1.86 \pm 0.16$, i.e. virtually identical to the 
red--selected LCRS.  
Clearly, given its fainter magnitude limit, in the ESP a much 
larger population of intrinsically faint, 
possibly less clustered, galaxies enters the sample.  This could
well be the reason for the smaller global clustering amplitude measured
from the magnitude-limited sample, where, as we explained above, faint
nearby galaxies play a dominant role on small scales in the J3--weighted 
technique.

Let us now look at \xip for the four "best" volume--limited samples,
whose plots are shown in Figure~\ref{csipz_vlim}.  The four panels display 
a number of interesting features.  First of all, one immediately
sees the striking compression of the contours at large
($>10\hmpc$) $r_p$'s for the two fainter, smaller-volume samples
in the top panels.  This effect is similar to what one would expect as 
consequence of infall motions onto superclusters, and if real could be
used to estimate again the parameter $\beta$ that
we discussed in the previous section (Kaiser 1987).  
A similar effect is not seen, however, in the two more luminous, deeper  
samples.  An explanation for this behaviour can be attempted by looking 
at the cone diagram of Figure~\ref{fig-cone}, and considering the limiting 
distances for the different volume--limited samples, 
(Table 1).  In the case of the $-18.5$ and $-19.0$ samples 
($z<0.11$ 
and $z<0.13$, respectively), one can see that two superclusters, nearly 
perpendicular to the line of sight, are dominating the galaxy distribution.  
This is clearly enhanced by their position near the far end of the sample, 
where the sampled volume is maximum.  In other words, more 
nearby, differently oriented, structures are just "cut-through" by the thin 
ESP slice, and weigh less, while these two are included nearly in their full 
size.   In this way they contribute  
an excess number of pairs along $r_p$, and thus an anisotropy
(compression) of the contours, which has nothing to do with a true
infall.  Note indeed how the effect is maximum for the $M\le-19$ sample, 
when both structures are fully included within the sample limit. 
Despite the depth of the ESP, therefore, these
two subsamples do not yet provide an unbiased (with respect to 
direction) set of superstructures.
Apparently, a more isotropic situation is reached within the 
two brighter samples, that cover a larger volume. In these
two cases, however, the signature of large--scale motions
is very weak.   An accurate study of this kind of distortions would  
require a careful modelling of all selection effects, and even with
this a measure of $\beta$ from the ESP looks rather difficult
(see e.g. Fisher \etal 1994b for details on the technique).

Still looking at Figure~\ref{csipz_vlim}, one can also notice that
the small--scale stretching of the contours is more pronounced for
the more luminous galaxies ($M\le-20$), than for the fainter ones.
Again, volume effects mix together with possible luminosity effects,
but one interpretation of this could be that luminous galaxies are 
more likely to be found in dense, high--velocity dispersion regions.
In fact, this is directly confirmed by our analysis of the
luminosity function of galaxies in ESP groups and clusters (Ramella et al.
1999), where we find that galaxies in high--density regions show a 
brighter $M^*$ with respect to ``field'' objects.   It is therefore
natural that selecting luminous galaxies, we also select galaxies
with higher pairwise dispersion.  

This picture is perfectly consistent with the behaviour of the real--space
correlation function for these samples.
Figure~\ref{wpfit_vlim} shows the results of the fit of a power--law
\xir to the projected function \wp, while Table~\ref{wpfit_tab}
summarises the corresponding best--fit values and errors.
Interestingly, in real space there is a weak, but systematic trend in both 
$r_o$ and $\gamma$ towards larger values
for larger luminosities. This is consistent with the higher small--scale
velocity dispersion inferred from the stretching of the contours of \xip.
One can also note from the left panels of Figure~\ref{wpfit_vlim}, that 
the effect is mostly produced on scales smaller than $2 \hmpc$, while no 
big change in amplitude is evident at larger separations.  Here we 
understand why no evident trend could be seen for the same four samples 
when we compared their redshift--space clustering through \xis: more 
luminous galaxies are indeed more clustered on small scales [higher \xir], 
but their pairwise dispersion increases accordingly and so the damping of 
\xis.   The net result is that in redshift space the two effects nearly 
cancel each other, and indeed nearly no sign of luminosity segregation is 
typically seen up to luminosities corresponding to $M\simeq -20$ (Benoist et al. 
1996).    

Similar real-space clustering analyses have been performed in the past on 
more "local" samples as the Perseus--Pisces (Guzzo et al. 1997), and 
the SSRS2 (Willmer et al. 1998) redshift surveys.  Even allowing
for systematic differences in the Zwicky and $b_J$ magnitude
systems, in general the ESP estimates tend to be smaller 
than those from samples of comparable 
mean luminosity extracted from these two surveys.  In particular,
values of the correlation length measured here for generic galaxies 
are comparable 
to those of spiral--only samples of similar luminosity from the 
Perseus--Pisces survey.
This would again suggest that the ESP is on the average richer in 
late-type galaxies.

\begin{table*}
\begin{center}
\begin{tabular}{lcc}  \hline\hline
Sample & $r_0\;(\hmpc)$ & $\gamma$ \\
\hline
mag--lim  &$4.15_{-0.21}^{+0.20}$ & $1.67_{-0.09}^{+0.07}$\\
--18.5&$3.45_{-0.30}^{+0.21}$ & $1.59_{-0.08}^{+0.07}$\\
--19&$4.05_{-0.22}^{+0.22}$ & $1.63_{-0.07}^{+0.06}$\\
--19.5&$4.25_{-0.27}^{+0.23}$ & $1.73_{-0.08}^{+0.08}$\\
--20&$5.15_{-0.44}^{+0.39}$ & $1.83_{-0.13}^{+0.11}$\\
\hline
\end{tabular}
\end{center}
\caption{Summary of the best--fit parameters of the real--space correlation 
function obtained from $w_p(r_p)$ for the different samples analysed.
}
\label{wpfit_tab}
\end{table*}                                                                   
%


\section{Tests for Potential Pathologies in the ESP}

The technical characteristics of the Optopus fibre spectrograph  
used for the redshift survey, introduced a bias at small angular separations, 
due to the finite size of the fibre connectors.  This forced a minimum 
separation of 25 arcsec, below 
which two objects could not be observed with the same plugged plate.  
Galaxies which are 
closer on the sky, therefore, had to be observed in two different exposures 
of the same Optopus field, to be able to collect a spectrum for both of them.
For some of the survey fields (the most crowded), these repeated observations
were not possible.  The consequence, explicitly shown in V98\footnote{Actually,
Figure 3 of that paper shows that the bias is
present out to separations of $\sim 50$ arcsec, indicating that in addition
to the mechanical limitation, there has also been some systematic tendency 
to avoid the very close pairs, when positioning the fibres during the
observations.}, 
is that when plotting the distribution of separations on the sky between each 
galaxy and its nearest neighbour, this is significantly more skewed towards 
small 
separations for the objects without redshift, than for the total photometric 
parent sample.  This could in principle introduce a bias into the correlation 
function estimate, and the consequences for \xis and \xip have therefore to be 
explored carefully. 

First, let us simply consider the fact that at the 
peak of the global redshift distribution function [$\propto z^2 \phi(z)$], 
around $z\sim 0.1$, the minimum fibre separation corresponds to less than 
$\sim 0.1 \hmpc$.  This should in principle affect directly 
only the first bin we used in our estimates of correlation functions: for 
$r$ much larger than the pair separation, i.e. the large scale bins, the 
proper pair count statistics should be guaranteed by the area-weighting 
scheme. 

To test any residual effect directly, we have made a further, simple exercise. 
Undoubtedly, the maximum bias on the correlation function occurs if
all the missing objects do really lie at the same distance of their nearest 
neighbour.  Let us therefore make the extreme assumption that all 
the close angular pairs with one missing redshift are true physical 
pairs, i.e. that $cz_1 \simeq cz_2$.    This situation should at least 
bracket the maximum bias we can expect from the missing pairs.  
To reproduce this case in a realistic fashion, we have selected all 
objects without redshift that are closer than 50 arcsec to a companion with 
measured redshift $cz_1$.  We have then assigned to each of these 
objects a fake 
redshift $cz_2 = cz_1+\delta_v$, where $\delta_v$ is a displacement extracted 
at random from a Gaussian distribution with $\sigma_{12}=350\kms$ 
\footnote{This value for $\sigma_{12}$ can be taken as a fiducial value 
for the pairwise dispersion in the field, as shown by Guzzo et al. (1997); 
note that in reality the line--of--sight distribution of pairwise 
velocities is best fitted by an exponential function, rather than 
a Gaussian.  This would however simply increase the tails of the 
simulated displacements: by using a Gaussian we are
placing the pairs a bit closer on the average, and therefore testing the 
effect on \x in an even more strict way.}.   Through this procedure, the
resulting sample with "measured" redshifts contains about 100
additional galaxies.  On this sample we repeated our computation of \xs
and \xip.  The $\j3$--weighted redshift--space correlation function is
compared in Figure~\ref{xis-test} to that from the original sample.
Evidently, the addition of the nearby pairs has very little effect,
indicating that the minimum--separation bias in the fibre positions
has no consequences on our correlation estimates.
Analogously, we re--computed \wp and obtained virtually no change in
the best fitting parameters for \xir.

\begf
\resizebox{\hsize}{!}{\includegraphics{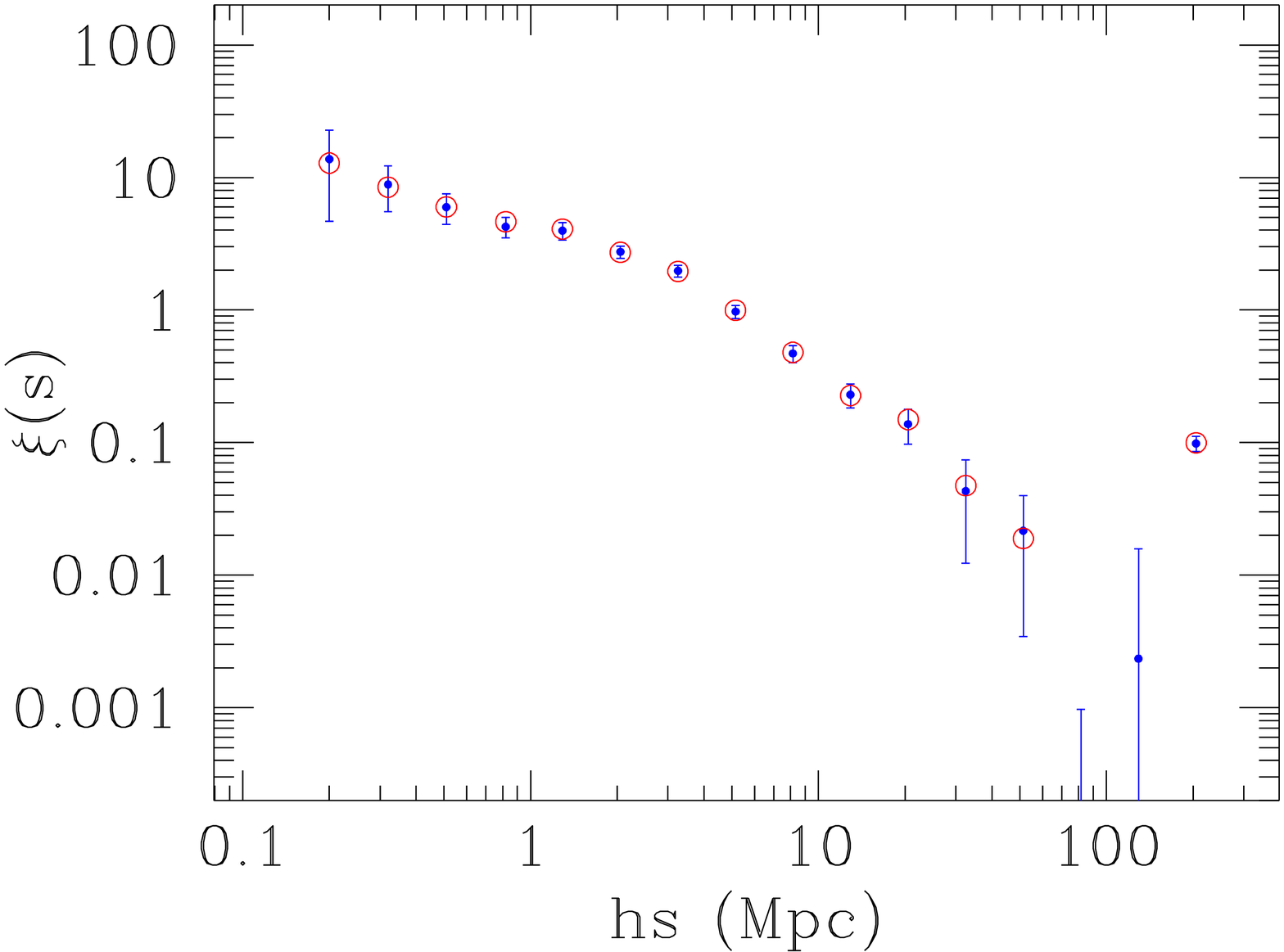}}
\caption{Test for the fibre distance bias. Open circles show the estimate 
of \xs from the modified sample where all close companions without 
redshift are assigned the redshift of the nearest neighbour 
(see text), compared to the estimate of \xis from the original catalogue.
Both estimates use the $J_3$ optimal weighting method.}
\label{xis-test}
\endf

\section{Summary}

The main results obtained in this work, can be summarised as follows.

\begin{enumerate}

\item The redshift-space correlation function \xis for the whole 
magnitude-limited ESP survey can be grossly described by a power law with 
$\gamma\sim 1.5$ between 3 and $\sim 50\hmpc$, where it smoothly 
breaks down, crossing the zero value between 60 and $80 \hmpc$.
However, if one considers the overall shape from 0.1 to $100\hmpc$,
a power law is a poor description of the observed \xis, and it is
not very meaningful to arbitrarily select one range or the other
for a formal fit.  In particular, below $2 \hmpc$, \xis has a much 
shallower slope ($\gamma < 1.$) .  This is shown to be mostly due to 
redshift-space depression by virialized structures, whose effect is 
enhanced by the $J_3$ optimal--weighting technique. 

\item Comparison to the only other redshift survey which provides a 
combination of depth and angular coverage similar or superior to ESP,
i.e. the LCRS, shows a very good agreement above $2\hmpc$.  
Minor differences are most easily explained in terms of the 
different selection procedures of the two galaxy samples.
Also comparison with a survey with very different geometry, depth
and redshift sampling (but with the same magnitude system), the Stromlo--APM,
shows a very similar behaviour for \xis.  The unanimity
of these surveys with rather different geometries and 
selection criteria, shows that the shape and amplitude of clustering 
for $\sim M^*$ optically--selected galaxies is now rather confidently
established, at least between 1 and $20\hmpc $.

\item Comparison of these redshift--space correlation functions to the best
available real--space correlation function from the APM survey, shows 
a negligible amplification from streaming motions, implying a low
value for the parameter $\beta = \Omega_o^{0.6}/b$.

\item Computing \xis from volume--limited subsamples of the ESP, we find 
no significant luminosity segregation in redshift space up to limiting 
absolute magnitudes as bright as $M_{b_J}=-20$.  We start seeing some
effect only for a subsample with $M\le-20.5$, for which the amplitude
of \xis increases by a factor $\sim 2$ on intermediate scales.  This
global behaviour in redshift space is in general agreement with what 
found by Benoist et al. (1996) on the SSRS2.  

These results also 
evidence how the small--scale shape of \xis as estimated using the
$\j3$--weighting technique is systematically flatter than that observed 
from any of the volume--limited samples, where no weighting is used.
Apparently, this is produced by enhanced weight put by the technique
on pairs of galaxies with high relative velocity (i.e. within
nearby clusters), resulting in a high
pairwise velocity dispersion and consequently in a stronger damping
of \xis below $2\hmpc$.   Although the effect is observed to be more
severe in the case of the ESP sample, probably because of its thin--slice
geometry [the LCRS \xis below $2\hmpc$ is closer to the volume--limited 
estimates], these differences underline the importance of comparing
weighted and unweighted estimates.

\item Studying clustering in real space through \xip and its projection \wp, 
we can evidence a mild dependence on luminosity also
for magnitudes fainter than $-20.5$.  This effect is mostly confined to
separations $r<2\hmpc$.  The shape of the contours of \xip also clearly 
shows how the small--scale pairwise velocity dispersion, and the 
corresponding distortions (mostly
produced within virialised structures), are increased in more luminous samples:
more luminous galaxies inhabit preferentially high--density regions,
where they consequently have to move faster with respect to each other.
This is the specific reason why no evident dependence
on luminosity is observed on small scales in redshift space, 
where the two effects cancel each other.  One could
speculate that this mild effect evidenced on small scales might be
the product of dynamical evolution of galaxies within virialized
regions, while that observed only for highly luminous galaxies on 
large scales, given the time scales implied, can only be produced
{\sl ab initio} in a biased process of galaxy formation.

\item Fitting \wp for the whole magnitude--limited ESP catalogue 
with a power--law \xir, we obtain as best parameters
$r_o=4.15^{+0.20}_{-0.21} h^{-1}\,$ Mpc and $\gamma=1.67^{+0.07}_{-0.09}$.
These amplitude and slope are slightly smaller than those from most other 
optical surveys, suggesting that late--type galaxies are probably slightly
favoured by the ESP selection, with respect to early types.

\end{enumerate}
 
\acknowledgements
LG is indebted to Karl Fisher for the use of his PCA fitting
routine and to Michael Strauss for several discussions on redshift--space
distortions.  This work has been partially supported through NATO Grant CRG 
920150, EEC Contract ERB--CHRX--CT92--0033, CNR Contract 95.01099.\-CT02 and
by Institut National des Sciences de l'Univers and Cosmology GDR.

\end{document}